\documentstyle[12pt,epsfig]{article}

\begin{document}

\begin{center}

{\large\bf
The present theoretical error on the Bhabha scattering cross section in 
the luminometry region at LEP}
\end{center}

\begin{center}
A.~Arbuzov$^1$, 
M.~Bigi$^2$, 
H.~Burkhardt$^3$, 
M.~Cacciari$^4$, 
M.~Caffo$^5$, 
H.~Czy\.z$^6$,
M.~Dallavalle$^7$, 
J.~H.~Field$^8$, 
F.~Filthaut$^9$, 
S.~Jadach$^{10}$,
F.~Jegerlehner$^{11}$, 
E.~Kuraev$^{12}$, 
G.~Montagna$^{13}$, 
O.~Nicrosini$^{14}$,
F.~Piccinini$^{15}$, 
B.~Pietrzyk$^{16}$, 
W.~P\l{}aczek$^{17}$, 
E.~Remiddi$^{18}$, 
M.~Skrzypek$^{19}$, 
L.~Trentadue$^{20}$, 
B.~F.~L.~Ward$^{21}$, 
Z.~W\c{a}s$^{22}$    
\end{center}

\vfil \hrule \footnotesize \noindent
$^1$BLTP, JINR, Dubna, 141980, Russia, ARBUZOV@THSUN1.JINR.DUBNA.SU; work 
supported in part by the Royal Swedish Academy of Sciences (ICFPM grant) \\ 
$^2$Dipartimento di Fisica Sperimentale, 
Universit\`a di Torino and INFN, Via P. Giuria 1, I-10125 Turin, Italy, 
MBIGI@VXCERN.CERN.CH \\
$^3$CERN, SL Division, CH-1211 Geneva 23, Switzerland, HBU@CERNVM.CERN.CH \\
$^4$DESY, Hamburg, Germany, CACCIARI@DESY.DE; DESY postdoc \\
$^5$INFN and Dipartimento di Fisica 
dell'Universit\`a, Bologna, Italy, CAFFO@BO.INFN.IT \\
$^6$Institute of Physics, University of Silesia, Katowice, Poland; 
INFN and Dipartimento di Fisica dell'Universit\`a, Bologna, Italy, 
CZYZ@BO.INFN.IT; work supported in part by
USA-Poland Maria Sklodowska-Curie Joint Fund II, grant MEN/NSF-93-145,
and Polish Committee for Scientific Research, grant no. 2P03B17708 \\
$^7$INFN and Dipartimento di Fisica 
dell'Universit\`a, Bologna, Italy, DVM@CERNVM.CERN.CH \\
$^8$University of Geneva, Switzerland, JFIELD@CERNVM.CERN.CH \\
$^9$Carnegie Mellon University, Pittsburgh, 
PA 15213, USA, FILTHAUT@HPL3.CERN.CH \\
$^{10}$Institute of Nuclear Physics, ul. Kawiory 26A, 30-055 Krak\'ow, Poland, and 
CERN, TH Division, CH-1211 Geneva 23, Switzerland, JADACH@CERNVM.CERN.CH; 
work supported in part by
Polish Committee for Scientific Research, grant no.~2P30225206,
US DoE contract DE-FG05-91ER40627,
and IN2P3 French-Polish Collaboration through LAPP Annecy \\
$^{11}$DESY Zeuthen, Germany and ETH Zurich, 
Switzerland, FJEGER@IFH.DE \\
$^{12}$BLTP, Dubna, 141980, Russia, KURAEV@THEOR.JINRC.DUBNA.SU \\
$^{13}$Dipartimento di Fisica Nucleare e Teorica and INFN, Pavia, Italy, 
MONTAGNA@PAVIA.PV.INFN.IT \\
$^{14}$INFN and Dipartimento di Fisica Nucleare 
e Teorica dell'Universit\`a, Pavia, Italy, NICROSINI@PAVIA.PV.INFN.IT \\
$^{15}$INFN, Pavia, Italy, PICCININI@PAVIA.PV.INFN.IT \\
$^{16}$LAPP, IN2P3-CNRS, F-74941 Annecy-le-Vieux Cedex, France, 
PIETRZYK@LAPP.IN2P3.FR \\
$^{17}$Department of Physics and Astronomy, 
The University of Tennessee, Knoxville, Tennessee 37996-1200, USA 
(On leave of absence from Institute of Computer Science, 
Jagellonian University, ul. Reymonta 4, Krak\'ow, Poland), 
PLACZEK@HEPHP02.PHYS.UTK.EDU; work supported in part by
US DoE contract DE-FG05-91ER40627 \\
$^{18}$INFN and Dipartimento di Fisica dell'Universit\`a, Bologna, Italy, 
REMIDDI@BO.INFN.IT \\
$^{19}$Institute of Nuclear Physics, 
ul. Kawiory 26A, 30-055 Krak\'ow, Poland, SKRZYPEK@HPJMIADY.IFJ.EDU.PL; 
work supported in part by
Polish Committee for Scientific Research, grant no.~2P30225206,
European Commission contract ERBCIPDCT940016,
and IN2P3 French-Polish Collaboration through LAPP Annecy \\
$^{20}$Dipartimento di Fisica, Universit\`a di Parma, and 
INFN, Gruppo Collegato di Parma, Sezione di Milano, Italy, 
TRENTA@CERNVM.CERN.CH, TRENTA@VXCERN.CERN.CH, TRENTADUE@ROMA2.INFN.IT \\
$^{21}$Department of Physics and Astronomy, 
The University of Tennessee, Knoxville, Tennessee 37996-1200, USA, and 
SLAC, Stanford University, Stanford, California 94309, USA, 
BFLW@SLACVM.SLAC.STANFORD.EDU; work supported in part by
US DoE contract DE-FG05-91ER40627 \\
$^{22}$CERN, TH Division, CH-1211 Geneva 23, Switzerland, and 
Institute of Nuclear Physics, ul. Kawiory 26A, 30-055 Krak\'ow, Poland, 
WASM@CERNVM.CERN.CH; work supported in part by
Polish Committee for Scientific Research, grant no.~2P30225206 \\ 

%%%%%%%%%%%%%%%%%%%%%%%%%%%%%%%%%%%%%%%%%%%%%%%%%%%%%%%%%%%%%%%%%%%%%
%%%%%%%%%%%%%%%%%%%%%%%%%%%%%%%%%%%%%%%%%%%%%%%%%%%%%%%%%%%%%%%%%%%%%
%%%%%%%%%%%%%%%%%%%%%%%%%%%%%%%%%%%%%%%%%%%%%%%%%%%%%%%%%%%%%%%%%%%%%
\vfil
\eject
\normalsize
\vfill
\begin{abstract}

{\small  The results concerning the theoretical evaluation of the 
small-angle
Bhabha Scattering cross section obtained during the  Workshop on Physics at
LEP2 (CERN, Geneva, Switzerland, 1995) by the Working Group ``Event Generators 
for Bhabha Scattering'' are summarized. The estimate of the theoretical
error on the cross section in the luminometry region is updated. }

\end{abstract}

\vfil
\begin{center}
Submitted to Physics Letters B
\end{center}

\vfil

%\leftline{\today}
\leftline{May 6, 1996}

\eject

%begin{figure}[hbtp]
%\begin{center}
%\epsfig{file=[montagna.paw]a2nl.eps,height=10truecm}
%\end{center}
%\caption{Comparison of MC's}
%\end{figure}

During 1995, within the Workshop on Physics at LEP2, a Working Group
on ``Event Generators for Bhabha Scattering'' was convened. The main 
tasks of the Working Group were

\begin{itemize}

\item to make an inventory of all the available Monte Carlo (MC) event
generators, developed by independent collaborations,
for small-angle (SABH) and large-angle (LABH) Bhabha processes at LEP1 and
LEP2;

\item to improve our understanding of their theoretical uncertainties by 
means of systematic comparisons of MC's
between themselves and with non-MC approaches.

\end{itemize}

The main emphasis was put on SABH processes, because of the pressing
need to match the theoretical precision of the calculations with the
much improved experimental accuracy ($\leq 0.1\%$) of the luminosity 
measurement. In
particular, the main achievement of the program outlined, which is the 
result
of a combined effort by several collaborations addressing several 
theoretical and
experimental issues, was the reduction of
the theoretical error on the SABH cross section from 0.16\% to 0.11\% 
for
typical event selections (ES) at LEP1, and a first estimate of the 
theoretical
error on the SABH cross section at LEP2.

The aim of the present short note is to summarize the strategies adopted 
in
order to achieve the goal stated above, and to officially state the 
conclusions
drawn by the Working Group. For any details concerning experimental 
aspects, or
 theoretical issues, as well as descriptions of the codes involved,
the reader is referred to~\cite{bharep}, where the proper references to 
all the
individual contributions can be found.

The various components of the theoretical error on the SABH cross section 
are
quoted in Tab.~\ref{tab:total-error-lep1}, where a summary of the past and
present situation at LEP1 together with the present estimate valid for 
LEP2
is given. The errors in the table are understood to be attributed
to the cross section for any typical (asymmetric) ES,
for a LEP1 experiment in the angular range $1^{\circ}-3^{\circ}$,
 calculated using BHLUMI~4.03.
In the case of LEP2, the estimate extends to the
  angular range $3^{\circ}-6^{\circ}$, and also to  a possible
  narrower angular range (say $4^{\circ}-6^{\circ}$) that may be
  necessary due to the effect of synchrotron radiation masks in the
  experiments.
The entries include combined technical and physical precision.

%%%%%%%%%%%%%%%%%%%%%%%%%%%%%%%%%%%%%%%%%%%%%%%%%%%%%%%%%%%%%%%%%%%%%%%%%%
%%%%%\begin{table}[!ht]
\begin{table}[hbtp]
\centering
\begin{tabular}{||l|l|l|l||}
\hline\hline
 & \multicolumn{2}{|c|}{LEP1} & LEP2 \\
\hline
  Type of correction/error
%%%%%%%%%& Ref. \protect\cite{th-95-38}
& Past
& Present
& Present     \\
\hline
(a) Missing photonic ${\cal O}(\alpha^2 L)$ &
    0.15\%      & 0.10\%    & 0.20\%
\\
(b) Missing photonic ${\cal O}(\alpha^3 L^3)$ &
    0.008\%     & 0.015\%    & 0.03\%
\\
(c) Vacuum polarization &
    0.05\%      & 0.04\%    & 0.10\%
\\
(d) Light pairs &
    0.01\%      & 0.03\%    & 0.05\%
\\
(e) Z-exchange  &
    0.03\%      & 0.015\%   &  0.0\%
\\
\hline
    Total  &
    0.16\%      & 0.11\%    & 0.25\%
\\
\hline\hline
\end{tabular}
\caption{\small\sf
Summary of the total (physical+technical) theoretical uncertainty
for a typical
calorimetric detector.
For LEP1, the above estimate is valid for the angular range
within   $1^{\circ}-3^{\circ}$, and
for  LEP2  it covers energies up to 176~GeV, and
angular range within $1^{\circ}-3^{\circ}$ and $3^{\circ}-6^{\circ}$
(see the text for further comments).
}
\label{tab:total-error-lep1}
\end{table}

\begin{figure}[hbtp]
\centering
%%%%%%%%%%%%%%%%%%%%%%%%%%%%\htmlimage{scale=1.6}
%---------------------------
\setlength{\unitlength}{0.1mm}
\begin{picture}(1700,1700)
%#####\put(0,0){\framebox( 1700,1700){ }}
%#####
\put( -20, 850){\makebox(0,0)[lb]{
\epsfig{file=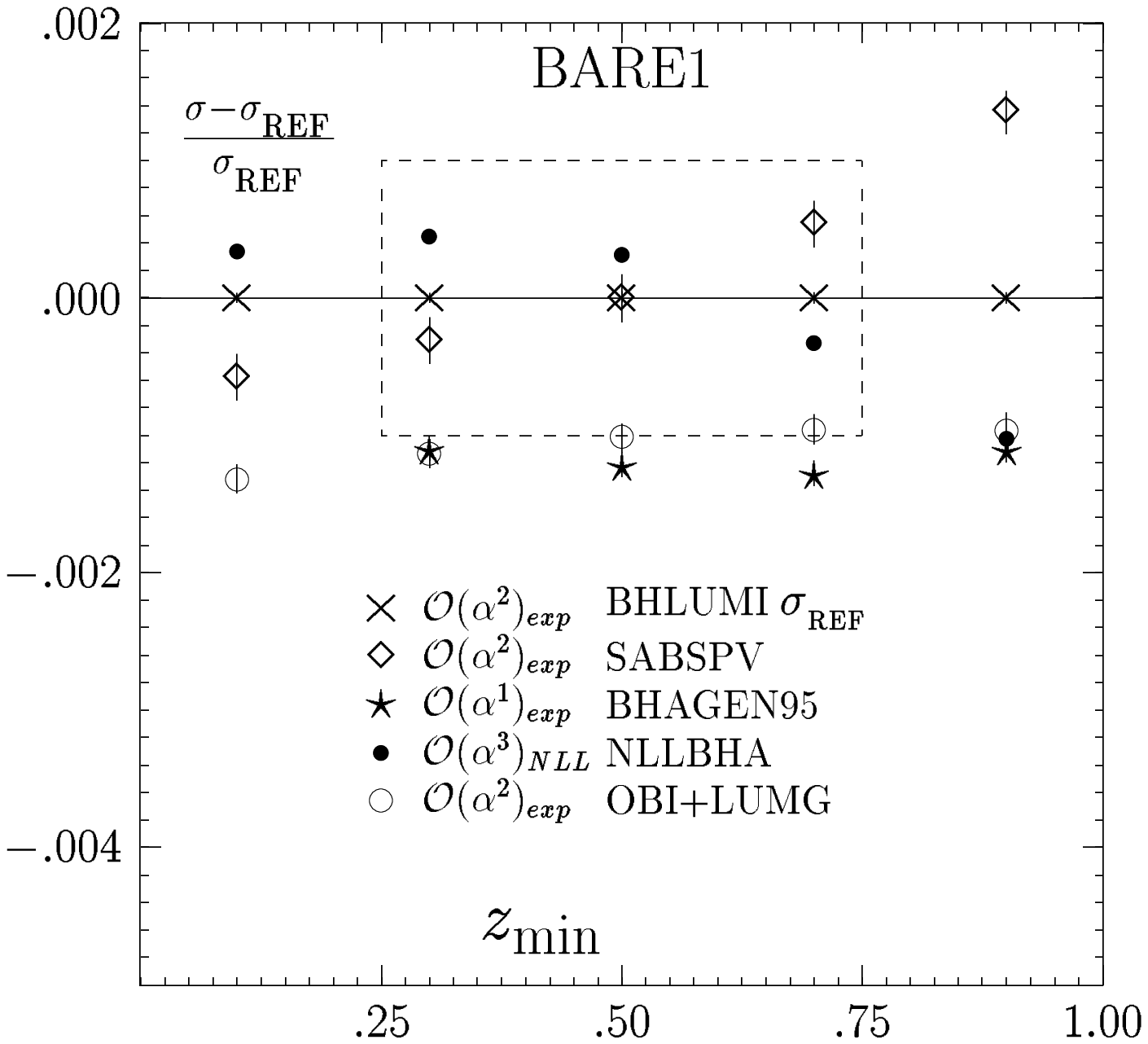,width=87mm,height=84mm}
}}
%#####
\put( 820, 850){\makebox(0,0)[lb]{
\epsfig{file=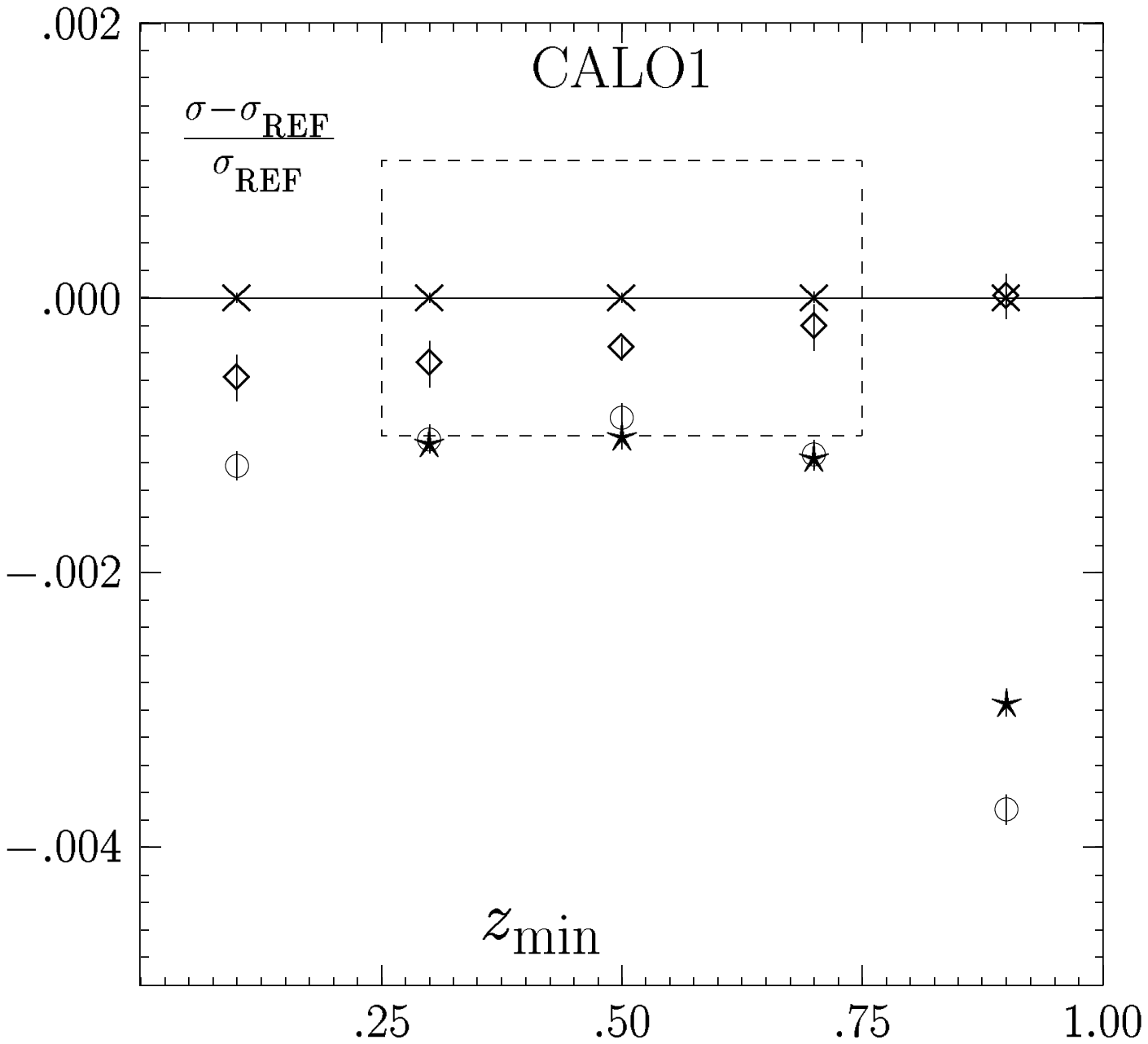,width=87mm,height=84mm}
}}
%#####
\put(-20,  00){\makebox(0,0)[lb]{
\epsfig{file=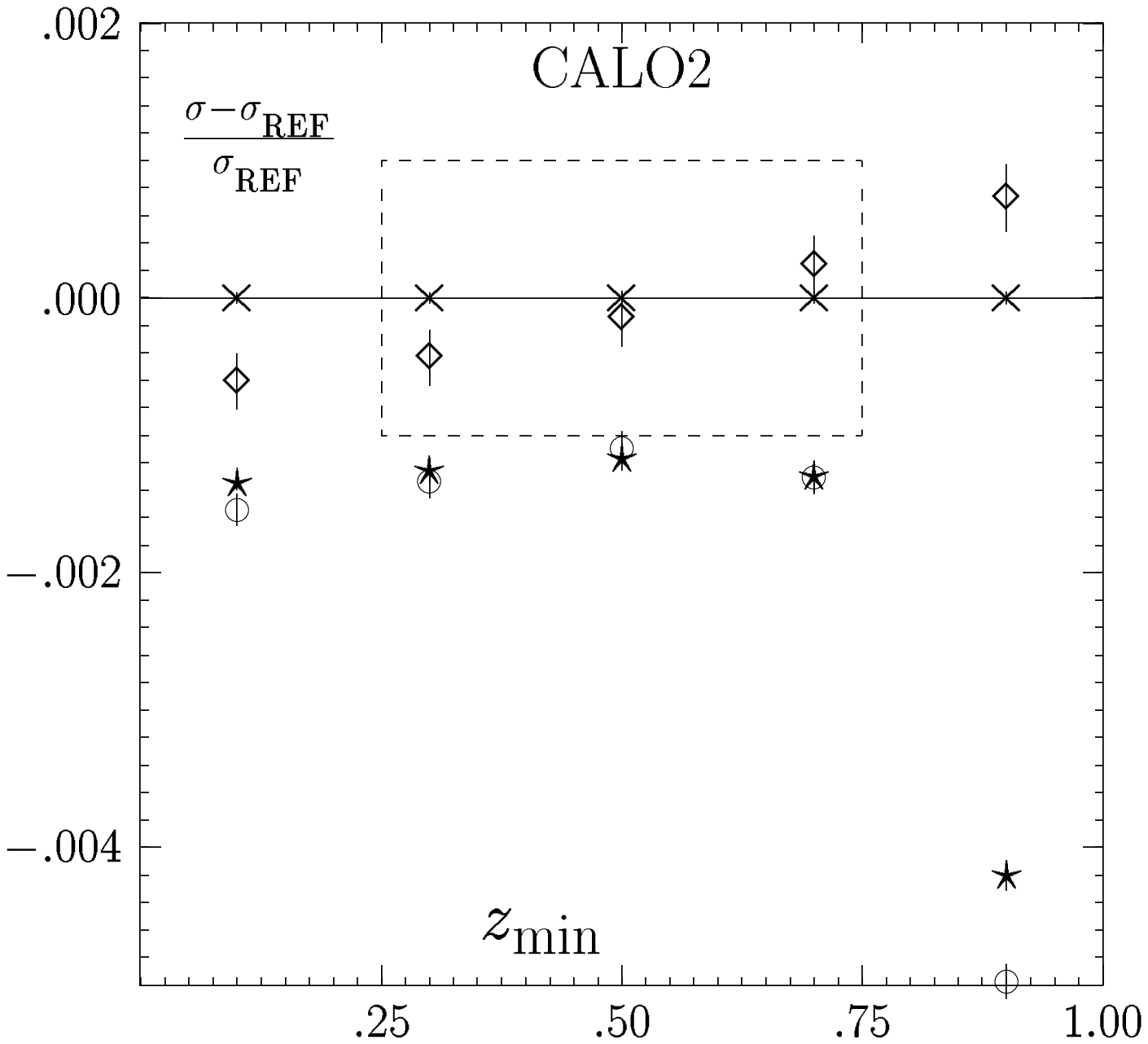,width=87mm,height=84mm}
}}
%#####
\put(820,  00){\makebox(0,0)[lb]{
\epsfig{file=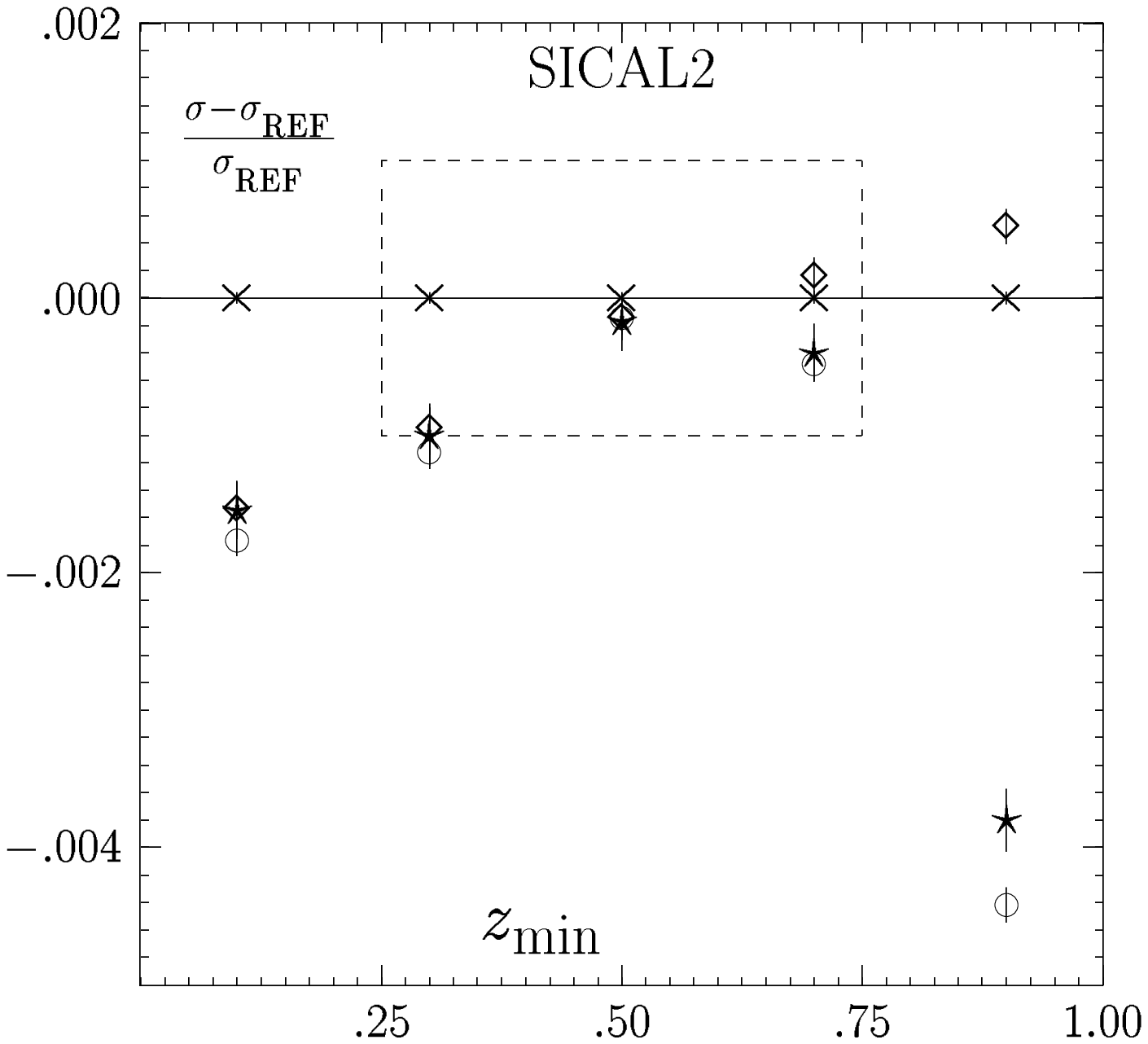,width=87mm,height=84mm}
}}
%%%%%%%%%%%%%%%%%%%%%%%%%%%%%%%%%%%%%%%%%%%%%%%%
%%% Various adds on the top of postscript %%%%%% 
%%%%%%%%%%%%%%%%%%%%%%%%%%%%%%%%%%%%%%%%%%%%%%%%
%\put(0,0){\begin{picture}( 1600,1600)
%  \put(  450,1630){\makebox(0,0)[b]{\small BARE1}}
%  \put( 1250,1630){\makebox(0,0)[b]{\small CALO1}}
%  \put(  450, 780){\makebox(0,0)[b]{\small CALO2}}
%  \put( 1250, 780){\makebox(0,0)[b]{\small SICAL2}}
%#####
%  \put( 250,1000){\begin{picture}( 300,300)
%    \put(  0,160){\makebox(0,0){\small $\times$}}
%    \put( 20,160){\makebox(0,0)[l]{\footnotesize
%        ${\cal O}(\alpha^2)_{exp}$}}
%    \put(180,160){\makebox(0,0)[l]{{\footnotesize BHLUMI}
%
%        {\footnotesize     $\sigma_{_{\rm{REF}}}$}  }}
%    \put(  0,120){\makebox(0,0){\small $\diamond$}}
%    \put( 20,120){\makebox(0,0)[l]{\footnotesize
%        ${\cal O}(\alpha^2)_{exp}$}}
%    \put(180,120){\makebox(0,0)[l]{\footnotesize SABSPV}}
%
%    \put(  0,80){\makebox(0,0){\small $\star$}}
%    \put( 20,80){\makebox(0,0)[l]{\footnotesize
%        ${\cal O}(\alpha^2)_{exp}$}}
%    \put(180,80){\makebox(0,0)[l]{\footnotesize BHAGEN94}}
%
%    \put(  0, 40){\circle*{15}}
%    \put( 20, 40){\makebox(0,0)[l]{\footnotesize
%        ${\cal O}(\alpha^3)_{_{NNL}}$}}
%    \put(180, 40){\makebox(0,0)[l]{\footnotesize NNLBHA}}
%
%    \put(  0,  0){\circle{20}}
%    \put( 20,  0){\makebox(0,0)[l]{\footnotesize
%        ${\cal O}(\alpha^2)_{exp}$}}
%    \put(180,  0){\makebox(0,0)[l]{\footnotesize OBS+LUMG}}
%  \end{picture}}
%#####
% \end{picture}}
\end{picture}
%%%%%%%%%%%%%%%%%%%%%%%%%%%%%%%%%%%%%%%%%%%%%%%%
%----------------------------
\caption{\small\sf
 Monte Carlo results for the symmetric Wide-Wide ES's 
 BARE1, CALO1, CALO2 and SICAL2, 
 for matrix elements beyond first order.
 Z exchange, up-down interference and vacuum polarization
 are switched off.
 The center of mass energy is $\protect\sqrt{s}=92.3$~GeV. 
 $z_{min}$ is the cut
 condition on the final-state energies, defined as $ E_+ E_- / E^2  \geq 
 z_{min}$, $E_{-,+}$ being the final-state energy of the bare electron and
 positron, respectively. The {\it fiducial } angular range is 0.024-0.058~rad
 (for more details on the
 ES, the reader is referred to~[1]).
 In the plot, the  ${\cal O}(\alpha^2)^{YFS}_{exp}$ cross section
% $\sigma_{_{\rm{BHL}}}$ 
from BHLUMI~4.03 is used as a reference cross section.
}
\label{fig:sical92-alf2exp}
\end{figure}

As can be seen in the table, at the present stage the theoretical error is
still dominated by the error on photonic corrections, quoted in entries 
(a) and
(b), and in particular by the one due to missing ${\cal O} (\alpha^2 L)$,
where $L$ is the usual collinear logarithm $L = \ln ( -t / m^2 )$, 
of entry
(a). Since the error of entry (a) is by far dominant with respect to all 
the
other ones, it is worth devoting some space to describe how it has been
estimated, namely by adopting  the following procedure:

\begin{itemize}

\item consideration of only the photonic corrections to the dominant
part of the SABH cross section, namely the one due to $t$-channel photon
exchange;

\item to define four families of ES's, starting from the simplest one, 
BARE1,
in which cuts are applied only to the ``bare'' final fermions, and going,
through CALO1 and CALO2, implementing more and more complex clustering
algorithms, to SICAL2, which is very similar to a ``real'' experimental 
ES;
since photonic corrections are very sensitive to the details of the ES,
defining these four ES's allows to span in detail the photonic phase 
space;
 even if the ES BARE1 is far from realistic,
the presently available
analytical calculation  including the complete set of
${\cal O} (\alpha^2 L)$ corrections refers to such an ES, and so
provides a very important cross-check of the MC programs;

\item to run all the available codes for all the ES's, varying inside any 
ES the
threshold requirements for the final state fermions/clusters;

\item to perform a test concerning the technical precision, namely 
comparing
the exact up to ${\cal O} (\alpha)$ cross sections; agreement at the 
level of a
few $10^{-4}$ relative deviation has been achieved (for brevity not 
shown here);

\item finally, to compare the results of the codes including the full 
higher-order
photonic corrections in each case, 
for all the situations explored (see Fig.~\ref{fig:sical92-alf2exp}).

\end{itemize}

\begin{figure}[!ht]
\centering
%%%%%%%%%%%%%%%%%%%%%%%%%%%%%%%%%%\htmlimage{scale=1.6}
%---------------------------
\setlength{\unitlength}{0.1mm}
\begin{picture}(1700,1700)
%#####\put(0,0){\framebox( 1700,1700){ }}
%#####
\put( 400, 850){\makebox(0,0)[lb]{
\epsfig{file=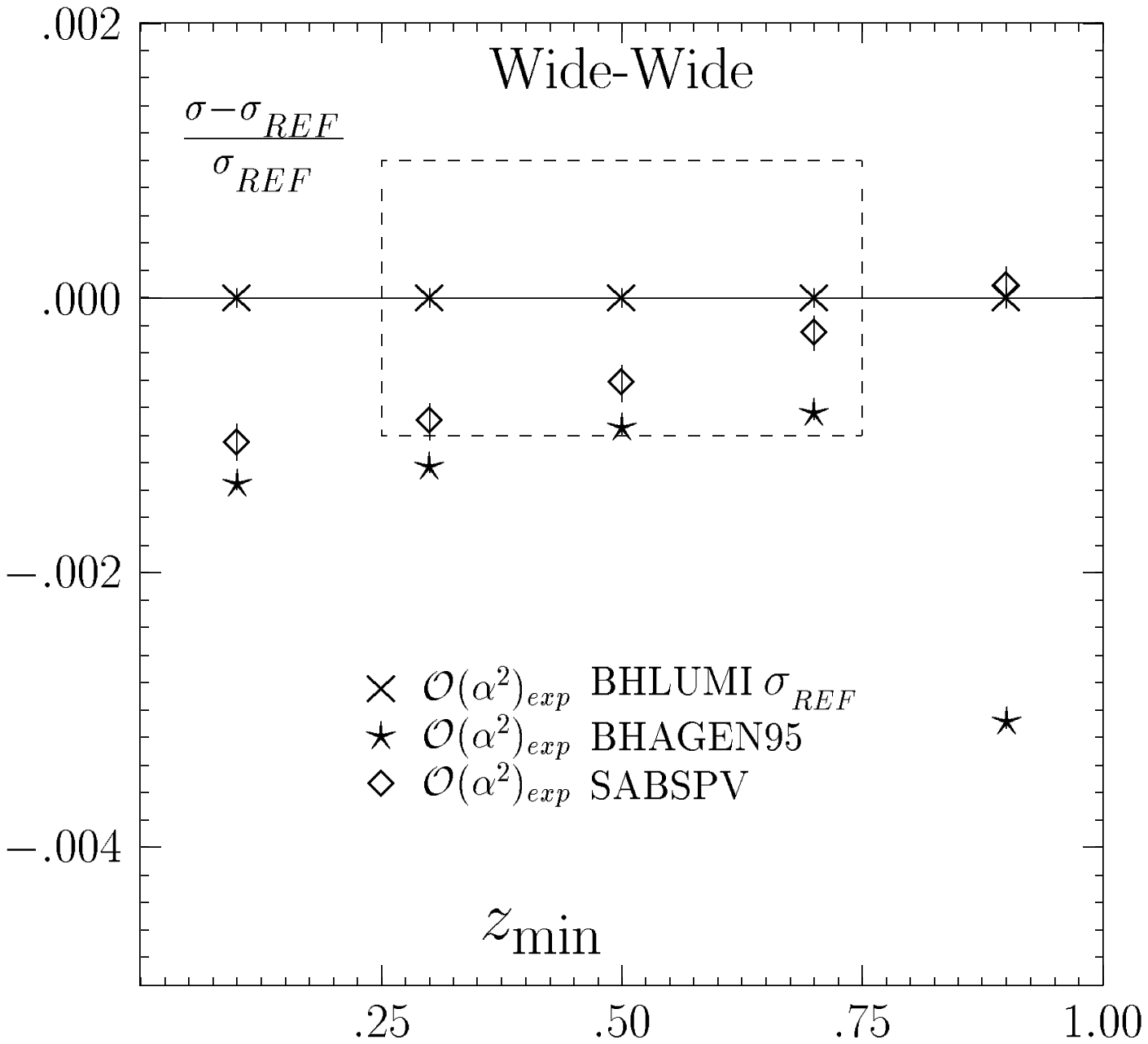,width=87mm,height=84mm}
}}
%##### 
\put(-20, 00){\makebox(0,0)[lb]{
\epsfig{file=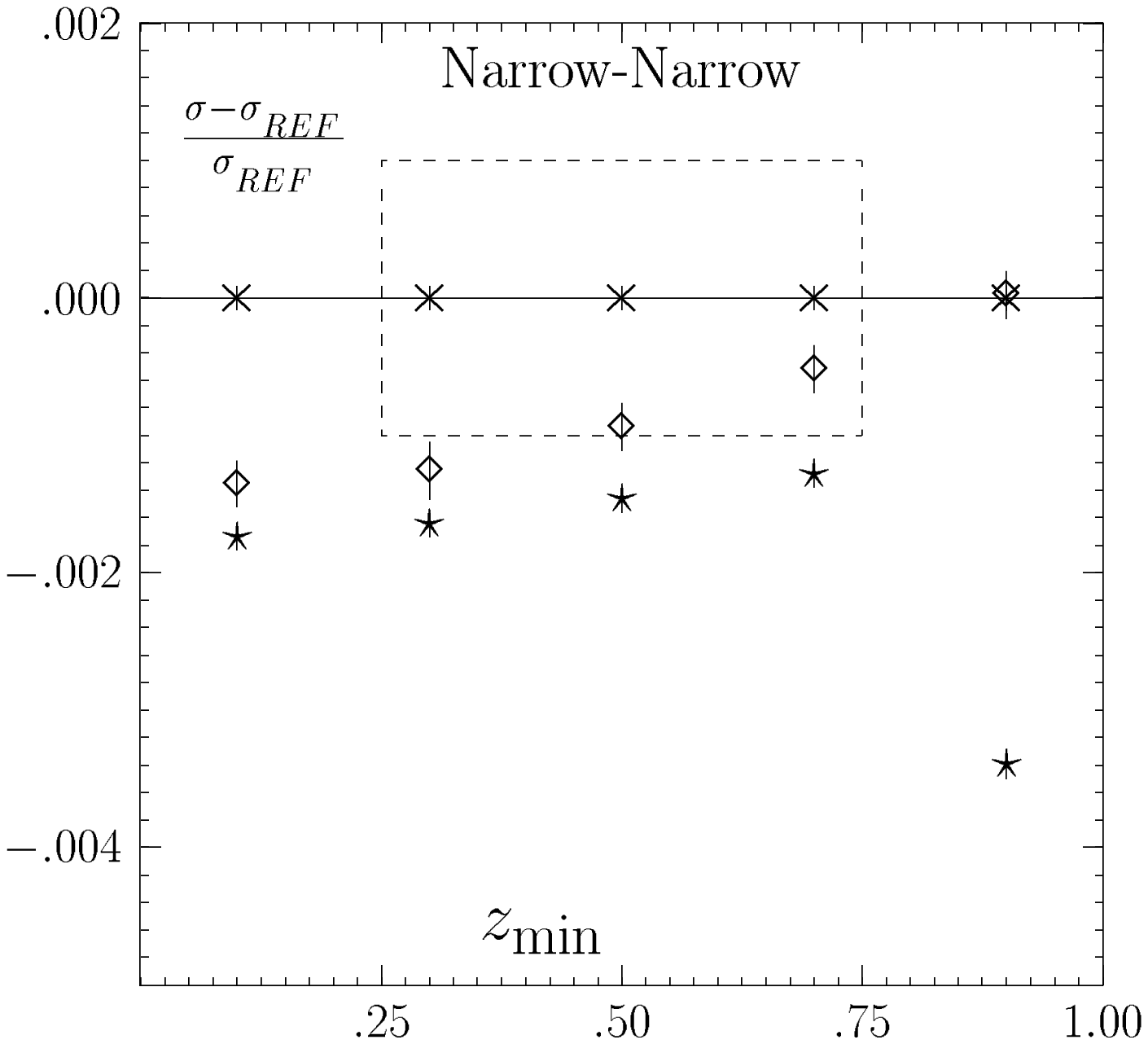,width=87mm,height=84mm}
}}
%#####
\put(820, 00){\makebox(0,0)[lb]{
\epsfig{file=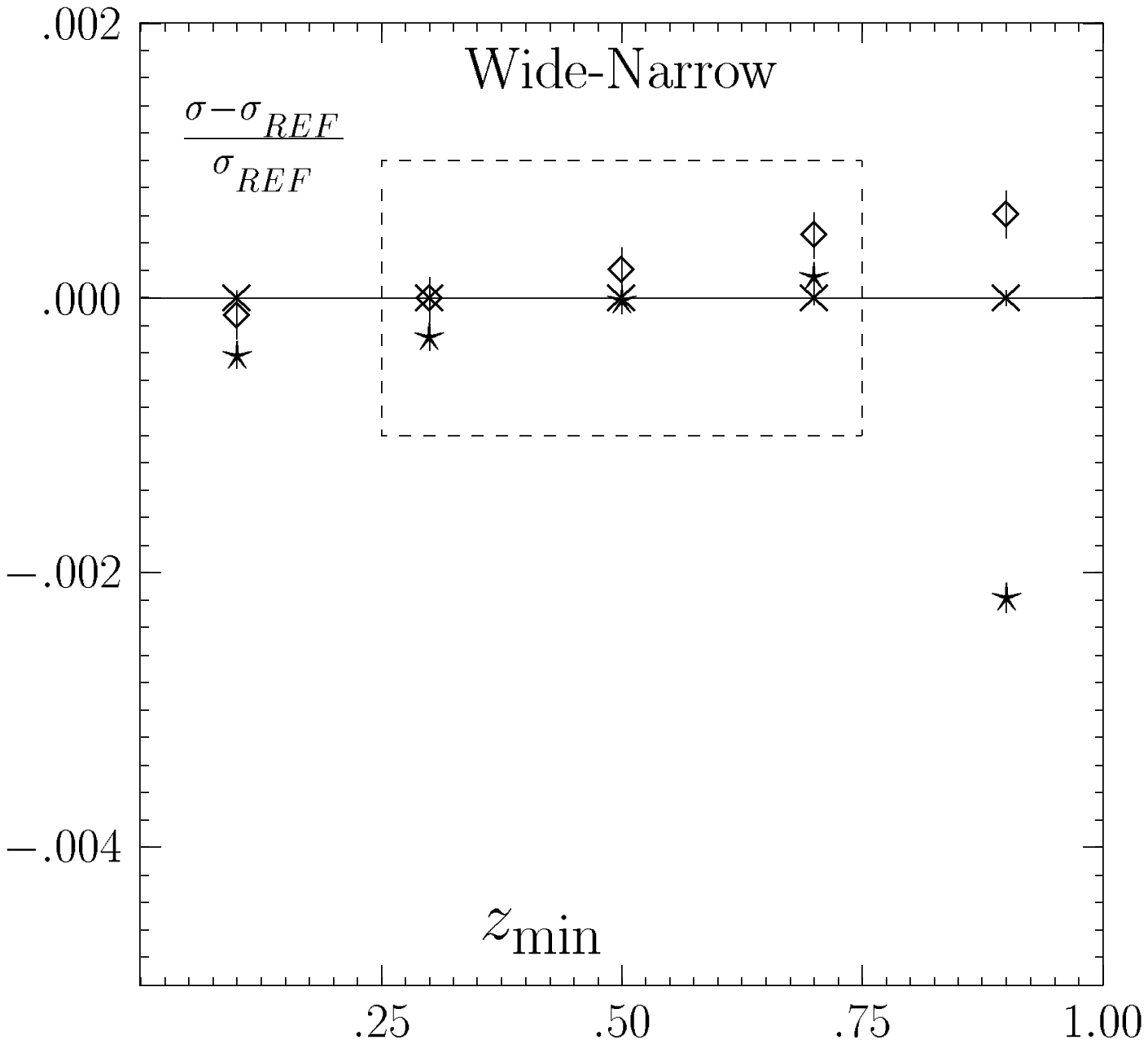,width=87mm,height=84mm}
}}
%%%%%%%%%%%%%%%%%%%%%%%%%%%%%%%%%%%%%%%%%%%%%%%%
%%% Various adds on the top of postscript %%%%%%
%%%%%%%%%%%%%%%%%%%%%%%%%%%%%%%%%%%%%%%%%%%%%%%%
\put(0,0){\begin{picture}( 1600,1600)
  \put(  200,1200){\makebox(0,0)[b]{\huge All}}
  \put( 1500,1200){\makebox(0,0)[b]{\huge Included}}
%  \put(  900,1630){\makebox(0,0)[b]{\small Wide-Wide}}
%  \put(  450, 780){\makebox(0,0)[b]{\small Narrow-Narrow}}
%  \put( 1300, 780){\makebox(0,0)[b]{\small Wide-Narrow}}
%%##### 
%  \put(700,1350){\begin{picture}( 300,300)
%    \put(  0,160){\makebox(0,0){\small $\times$}}
%    \put( 20,160){\makebox(0,0)[l]{\footnotesize 
%        ${\cal O}(\alpha^2)_{exp}$}}
%    \put(180,160){\makebox(0,0)[l]{{\footnotesize BHLUMI} 
%        {\footnotesize     $\sigma_{_{\rm{REF}}}$}  }}
%    \put(  0,120){\makebox(0,0){\small $\diamond$}}
%    \put( 20,120){\makebox(0,0)[l]{\footnotesize 
%        ${\cal O}(\alpha^2)_{exp}$}}
%    \put(180,120){\makebox(0,0)[l]{\footnotesize SABSPV}}
%%    \put(  0,80){\makebox(0,0){\small $\star$}}
%%    \put( 20,80){\makebox(0,0)[l]{\footnotesize 
%%        ${\cal O}(\alpha^2)_{exp}$}}
%%    \put(180,80){\makebox(0,0)[l]{\footnotesize BHAGEN94}}
%%    \put(  0, 40){\circle*{15}}
%%    \put( 20, 40){\makebox(0,0)[l]{\footnotesize 
%%        ${\cal O}(\alpha^2)_{exp}$}}
%%    \put(180, 40){\makebox(0,0)[l]{\footnotesize OBS+LUMG}}
%  \end{picture}}
%%#####
\end{picture}} 
\end{picture}
%%%%%%%%%%%%%%%%%%%%%%%%%%%%%%%%%%%%%%%%%%%%%%%%
\caption{\small\sf
 Monte Carlo results for various symmetric/asymmetric           
 versions of the CALO2 ES,
 for matrix elements beyond first order.                    
 Z exchange, up-down interference and vacuum polarization       
 are switched ON.                                               
 The center of mass energy is $\protect\sqrt{s}=92.3$~GeV.
 $z_{min}$ is the cut
 condition on the final-state energies, defined as $ E_+ E_- / E^2 \geq 
 z_{min}$, $E_{-,+}$ being the final-state energy of the electron and
 positron clusters, respectively. 
 The {\it fiducial } angular range is 0.024-0.058~rad (for more details on the
 ES, the reader is referred to~[1]). 
% Not available x-sections are set to zero.                          
 In the plot, the ${\cal O}(\alpha^2)^{YFS}_{exp}$ cross section     
% $\sigma_{_{\rm{BHL}}}$ 
from BHLUMI~4.03 is used as a reference cross section.                          
% end-of-caption                                                
}
\label{fig:sical92asy-FG}
\end{figure}

The result of this procedure allowed the definition of ``one-per-mill 
regions'',
referring to realistic threshold cuts, within which most of the 
predictions lie.
Moreover, for those cases for which the predictions do not lie within the
``one-per-mill regions'', the reasons for the deviations involved have 
been
carefully investigated and eventually understood.
An analogous procedure has been followed after the inclusion of all the
relevant radiative corrections (vacuum polarization, $Z$ contributions 
and so
on), and extending the comparisons also to asymmetric ES's,
leading to the results shown as an example in 
Fig.~\ref{fig:sical92asy-FG}.\footnote{\footnotesize 
The results by BHAGEN95
shown here are slightly changed with respect to the ones quoted
in~\cite{bharep}, due to a bug-fixing. }
A similar
analysis has also been performed for the first time in situations which 
will be
typical at the LEP2 experiments.
The conclusion drawn at the end of all these comparisons is that now the
theoretical uncertainty due to uncontrolled ${\cal O} (\alpha^2 L)$ 
corrections
is reduced from 0.15\% to 0.10\% for the LEP1 situation, and estimated to be
0.20\% at LEP2. As far as entry (b), the 
``missing photonic $\cal O (\alpha^3 L^3) $''
uncertainty, is concerned, new estimates of the effect have resulted in a more
conservative theoretical error, namely 0.015\% to be compared with the old
estimate of 0.008\%.\footnote{\footnotesize In the semi-analitical 
program NLLBHA in the BARE1 set-up of Fig.~1
  all the corrections of $\cal O (\alpha^2 L)$ and of $ \cal O (\alpha^3 L^3) $
   are contained.}

As far as all the other entries in Tab.~\ref{tab:total-error-lep1} are
concerned, namely entries (c), the ``vacuum polarization'' uncertainty, 
(d),
the ``light pairs'' uncertainty, and (e), the ``$Z$-exchange'' 
uncertainty,
two of them, (c) and (e), 
are reduced with respect to the previous situation
thanks to 
several new fits
of the hadronic contribution to the vacuum polarization,
and some
additional original work on the $Z$-exchange contribution done during the 
workshop. New estimates, both Monte Carlo and analytical, of
the light pairs contribution, (d), featuring more complete calculations
done during the workshop, have resulted in a more conservative estimate
of the pairs effect uncertainty of 0.03\%;\footnote{\footnotesize Table~20 
in~\cite{bharep} contains a copying error in its light pairs row  
for the Past LEP1 entry, where 0.04\% should be 0.01\%, as we show it here.} 
if it would be necessary, 
this effect can be included in the BHLUMI event simulation itself
using an extension of the respective
YFS exponentiation to soft pairs radiation
via methods already represented in~\cite{bharep} and references therein.

In conclusion, the total (physical + technical) theoretical uncertainty 
on the
SABH cross section for a typical calorimetric detector is at present 
0.11\% at
LEP1 and 0.25\% at LEP2.

{\bf Acknowledgements --} M.~Cacciari thanks the Fondazione A.~Della Riccia 
and the Universit\`a di Pavia for supporting his staying at DESY when 
this work was performed.
H.~Czy\.z thanks OPAL collaboration for the support and kind hospitality
  during completion of this work. 
B.F.L.~Ward thanks Profs. G.~Veneziano and G.~Altarelli
and Prof. D.~Schlatter for the support
and kind hospitality of the CERN TH Division and of the
ALEPH Collaboration, respectively, while this work
was completed.

%%%%%%%%%%%%%%%%%%%%%%%%%%%%%%%%%%%%%%%%%%%%%%%%%%%%%%%%%%%%%%%%%%%%%%%%%%%

%%%\end{document}
%%%%%%%%%%%%%%%%%%%%%%%%%%%%%%%%%%%%%%%%%%%%%%%%%%%%%%%%%%%%%%%%%%%%%%%%%
\newpage
%
%   ######################################
%   ##  BEGIN   sical92-alf2exp.txi    ###
%   ######################################
%

%%%%%%%%%%%%%%%%%%%%%%%%%%%%%%%%%%%%%%%%%%%%%%%%%%%%%%%%%%%%%%%
%%%%   make sical92-alf2exp.dvi
%%%%%%%%%%%%%%%%%%%%%%%%%%%%%%%%%%%%%%%%%%%%%%%%%%%%%%%%%%%%%%%

%\begin{table}[!ht]
%\centering
%%%%%%%%%%%%%%%%%%%%%%%\htmlimage{scale=1.4}
%\setlength{\unitlength}{0.1mm}
%#####
%\begin{picture}(1700,1300)
%#####\put(0,0){\framebox( 1700,1300){ }}
%#####
%\put( 20, 00){\makebox(0,0)[lb]{
%\epsfig{file=sical92-alf2exp-Tab.ps,height=130mm,width=165mm}
%}}
%#####
%\end{picture}
%#####
%\caption{\small\sf
% Monte Carlo results for the symmetric Wide-Wide ES's 
% BARE1, CALO1, CALO2 and SICAL2, 
% for matrix elements beyond first order.
% Z exchange, up-down interference and vacuum polarization
% are switched off.
% The center of mass energy is $\protect\sqrt{s}=92.3$~GeV.
% Not available x-sections are set to zero.
%}
%\label{tab:sical92-alf2exp}
%\end{table}

%%%%%%%%%%%%%%%%%%%%%%%%%%%%%%%%%%%%%%%%%%%%%%%%%%%%%%%%%%%%%%%
%%%%%%%%%%%%%%%%%%%%%%%%%%lframe%%%%%%%%%%%%%%%%%%%%%%%%%%%%%%%
%%%\begin{figure}[!ht]
%------------------------------------------------------------------------------
%------------------------------------------------------------------------------
%------------------------------------------------------------------------------

%
%   #####################################
%   ##  END    sical92-alf2exp.txi    ###
%   #####################################
%

%%%\end{document}
%%%%%%%%%%%%%%%%%%%%%%%%%%%%%%%%%%%%%%%%%%%%%%%%%%%%%%%%%%%%%%%%%%%%%%%%%%%%
\newpage

%
%   ######################################
%   ##  END      sical92asy-FG.txi     ###
%   ######################################
%

\end{document}